\documentclass[aps,prl,showpacs,twocolumn,groupedaddress,floatfix]{revtex4}
\usepackage{amsmath,bm}
\usepackage{mathrsfs}
\usepackage{amsfonts}

\usepackage{graphicx}
\usepackage{xcolor}
\usepackage{ulem}

\makeatletter
\newcommand\colorsout[1]{\bgroup \markoverwith{\textcolor{#1}{\rule[0.5ex]{2pt}{0.4pt}}}\ULon}

\makeatother

\begin{document}

\title{Spontaneous spin switching via substrate-induced decoherence}

\author{Jean-Pierre Gauyacq}
\affiliation{Institut des Sciences Mol\'eculaires d'Orsay, ISMO,
 Unit\'e mixte CNRS-Univ
 Paris-Sud,UMR  8214, B\^atiment 351, Univ
 Paris-Sud, 91405 Orsay CEDEX,
France}
\author{Nicol\'as Lorente}
\affiliation{
 ICN2 - Institut Catala de Nanociencia i Nanotecnologia, Campus UAB, 08193 Bellaterra (Barcelona), Spain \\
CSIC - Consejo Superior de Investigaciones Cientificas, ICN2 Building, 08193 Bellaterra (Barcelona), Spain
}

\date{\today}
\begin{abstract}
We analyze the spontaneous switching between the two N\'eel states
of  Fe chains  on Cu$_2$N/Cu(100), experimentally studied  by Loth
et al. [Science \textbf{335}, 196 (2012)].  We show that, in the
experimental low-temperature regime ($T \approx$ 1~K),  decoherence
induced by substrate electrons deeply modifies the  dynamics of the
chain magnetization.  The Rabi oscillations associated to quantum
 tunneling of the isolated chain are replaced by an irreversible decay
 of considerably longer characteristic time.  The computed switching
rate  is small, rapidly decreasing with chain length, with  a $1/T$
behavior and in good agreement with the experiment. Quantum tunneling is
only recovered as the switching mechanism at extremely low temperatures,
the corresponding cross-over moving rapidly to further low temperatures
as the chain length is increased.
\end{abstract} 
\maketitle

Decoherence is becoming a central concept in the study of
quantum objects~\cite{Zurek1991,Zurek2003,Schlosshauer2004}.
It is caused by the influence of the environment on the studied
object that leads to the suppression of interference between
quantum states and hence to {a classical behavior
of the  object.} The presence of decoherence is
ubiquituous: from atomic optics~\cite{Haroche,Jacquod,Eiguren}, to the
foundations of quantum mechanics and the passage to the classical
world~\cite{Zurek1991,Zurek2003,Schlosshauer2004}, including
the environment-imposed limitations in quantum information and
computations~\cite{DiVicenzo,FernandoLuis}.  Decoherence also plays an
important role in surface science~\cite{Doyen2009}, like in the scattering
and localization of  atoms on surfaces, in the {dynamics}
of surface-confined electronic states~\cite{Hoefer,Olsson} or in
surface-adsorbed spins~\cite{Doyen2009,FR2012}. For an
adsorbate, the substrate acts as a bath of electronic, phononic
and nuclear excitations that cannot be neglected in the study of
adsorption. The scanning
tunneling microscope (STM) has permitted us to have unprecedented
insight on adsorbed quantum objects. Indeed, the STM has been
used to unveil the static and dynamical properties of spins on
surfaces~\cite{Heinrich2004,Hirjibehedin2007,Loth2010,Hambourg}. In
this context, decoherence plays a particular important
role~\cite{FR2012,FR2014,FR_private}.

Recently, Loth and collaborators~\cite{Loth2012} measured the dynamical
properties of antiferromagnetically coupled atomic Fe chains on
Cu$_2$N/Cu (100).  They oberved a N\'eel spin ordering
in chains with an even number of atoms, i.e. all the spins
are aligned on the same direction with alternate directions along the
chain. Their experiment revealed  that the chains could 
{switch} from one N\'eel state to the opposite N\'eel state
via either electron injection from an STM tip or
by increasing the sample temperature.  The tunneling
electron-induced regime showed a clear threshold when the electron energy
was large enough to excite an intermediate magnetic excited state of the
chain that greatly enhanced the switching~\cite{Loth2012,Gauyacq2013a}.
When studying the temperature-induced flipping of the local
moments of the chains,~\cite{Loth2012} two different regimes were
found: ($i$) a high-temperature
regime where the switching rate followed {an Arrhenius
exponential behavior with a finite activation energy that is in good
agreement with the energy of the intermediate state of the tunneling
electron-induced process} and ($ii$) a low-temperature regime, flat as a
function of inverse of temperature, reminiscent of the customary quantum
tunneling regimes of atom diffusion~\cite{Goran,Lauhon} or of magnetic
processes~\cite{Barbara1990,Prokofev,QT,QT1}. Regime $(i)$ {is analogous
to the STM electron-driven process and it} has been carefully analyzed
in the literature~\cite{Loth2012,Gauyacq2013a}, however, regime $(ii)$
has received less attention due to the conceptual difficulties that it
presents. Indeed, we show here that the {very low-switching
rate of regime $(ii)$ and its weak temperature dependence are} not due to
a tunneling process~\cite{Loth2012}, 
but {are rather  an effect of the fast substrate-induced
decoherence;  this} places decoherence at the heart of spin stability
in {a few-atom devices}.

\begin{figure}
\includegraphics[width=0.45\textwidth]{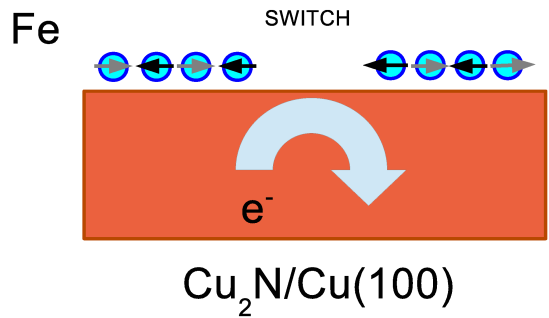}
\caption{
Switch between the two N\'eel states of an Fe atomic chain on a Cu$_2$N/Cu
(100) substrate. The thermal electrons {from the substrate}
cause the decoherence of the chain's spin state. 
{The joint action of the high-order indirect
Heisenberg coupling between the N\'eel states and of the substrate-induced decoherence results in the 
spontaneous switching of the chain between the two magnetization states.}
 \label{figure1}
}
\end{figure}

Using the theory developed in Refs.~\cite{Lorente2009,PSS}, we have
evaluated the switching rate between the above N\'eel states.  Only when
decoherence is included, can the experimental data
be explained.  Let us suppose that initially, the Fe
chain is prepared in one of the two N\'eel states. In the absence of
decoherence, a Rabi oscillation sets in between the two N\'eel states,
because the N\'eel states are broken-symmetry solutions of the chain's
Hamiltonian~\cite{Gauyacq2013a}. The period of the Rabi oscillation is
proportional to the inverse of the energy separation
between the two lowest-lying eigenstates of the chain, following the
customary quantum-tunneling picture~\cite{QT,QT1}.
These oscillations {are} many  orders of magnitude faster
than the measured switching rates~\cite{Loth2012}. However, when
decoherence induced by {thermal} electron-hole pair excitations of the substrate
{is} considered, Fig.~\ref{figure1}, the oscillations are damped {and
{the} decoherence} produces the collapse of the spin state into one of the
two N\'eel states with equal probability.

The Fe chain {is described by a set of local spins, $\vec{S}_i$, {in} an anisotropic environment and coupled by a first-neighbor Heisenberg exchange, $J$,}~\cite{Heinrich2006,Fransson2009,FR2009,Loth2012,Gauyacq2013a,PRB2011}:
\begin{eqnarray}
\hat{H}&=&\sum_{i=1}^{N-1}  ( J \vec{S}_i \cdot \vec{S}_{i+1} +
g \mu_B \vec{S} \cdot \vec{B} \nonumber \\
&+&
D S_{i,z}^2 +
E ( S_{i,x}^2 - S_{i,y}^2)  ).
\label{hamiltonien}
\end{eqnarray}
The local Fe spins are $S_i=2$, {where $i$ is the atom index for $N$ Fe atoms}. Local axial, $D$, and
transverse, $E$, magnetic anisotropies have been included together with
the Zeeman term due to an external field $\vec{B}$.  The actual values
of $J$, $D$, $E$ and $g$ are taken from the study {on
adsorbed dimers} by Bryant \textit{et al}~\cite{Bryant2013}.  Due to the
{above interactions and the} integer total spin, the ground state, $|GS\rangle$, is not degenerate. 
{In systems like the present one with a large D anisotropy term, the system is close to an Ising model.}
{The}
first excited state, $|EXC\rangle$, has a very small excitation
energy {that we express as} $2V$. 
Both $|GS\rangle$ and $|EXC\rangle$ correspond to the
superposition of many configurations of local spins, with  large weights of equal value on the two N\'eel
states. From these, one can then define the N\'eel states, $|N_1\rangle$
and $|N_2\rangle$, as linear combinations of the two low lying states:  
\begin{eqnarray}
|N_1\rangle &=& \frac{1}{\sqrt{2}} ( |GS\rangle+|EXC\rangle), \nonumber \\
 |N_2\rangle &=& \frac{1}{\sqrt{2}} ( |GS\rangle-|EXC\rangle).
\label{GS}
\end{eqnarray}
With this definition, the two N\'eel states are not
pure N\'eel configurations of local spins, but include
correlation effects that are important in {anti-ferromagnetic
chains}~\cite{desCloizeaux,Gauyacq2013a,PRB2011}. {The above
excitation energy, $2V$, then appears as twice} the coupling
{between} the two N\'eel states, Eq.~(\ref{GS}).  {The coupling}
$V$ is very weak {and} it is due to the correlated nature of the
N\'eel states. {Actually, it is possible to go from $|N_1\rangle$ to
$|N_2\rangle$ by applying several times the {Heisenberg
exchange  {or the anisotropy $E$ operators in} Eq.~(\ref{hamiltonien})}. The two
states are then weakly coupled via a high-order indirect interaction
involving many intermediate spin configurations.} {Due to this
high-order character, the coupling $V$ is very sensitive to the choice
of parametrization in Eq.~(\ref{hamiltonien}).}

When the chain is prepared in the N\'eel state $|N_1\rangle$, the chain
evolves under the above effective coupling, $V$,
leading to Rabi oscillations between the two states $|N_1\rangle$ and
$|N_2\rangle$ with period $T_{Rabi}=\pi \hbar/V$. This {corresponds to a quantum tunneling phenomenon and }
would account for the {periodic} spontaneous switching of magnetization of the chain if
it were isolated.

The effect of the substrate can be included using the density matrix of the system, 
$\hat{\rho}_T=|\Psi\rangle\langle \Psi|$
where $|\Psi \rangle$ is the state of the full system.
By tracing out the environment's degrees of freedom, we are left with the reduced
density matrix for the spin degrees of freedom of the chain~\cite{cohen}: 
\begin{equation}
\hat{\rho}=\sum_{j} \langle j|_{env} (|\Psi\rangle\langle \Psi|) |j\rangle_{env},
\end{equation}
where $|j\rangle_{env}$ is a complete basis set of the environment
system, $env$.  The time evolution of the reduced density matrix 
is then given by:
\begin{equation}
i \hbar \frac{d \hat{\rho}}{d t} = [{\hat{H}}_{red}, \hat{\rho}] + \hat{R} (\hat{\rho}),
\label{pilot}
\end{equation}
where ${\hat{H}}_{red}$ is  the 
$2\times2$ Hamiltonian matrix for the space spanned by
the two N\'eel states (the two diagonal energies are equal and the
two non-diagonal terms are equal to the effective coupling, $V$).
$\hat{R}$ takes into account the effect of the environment in the
reduced density matrix evolution, it is given by the
expresion~\cite{cohen,Molmer}:
\begin{equation}
\hat{R} (\hat{\rho}) = -i \Gamma \, ( \,|N_1\rangle \rho_{12} \langle N_2 | +
|N_2\rangle \rho_{21} \langle N_1 | \,),
\label{R}
\end{equation}
where $\rho_{12}$ and $\rho_{21}$ are the non-diagonal density matrix elements between the two
N\'eel states, $|N_1\rangle$ and $|N_2\rangle$.  
The {quantity} $\Gamma/\hbar$ is the decoherence rate of a N\'eel state,
{given by the} inverse of the pure dephasing time, $T_2^*$.
The substrate is  not   magnetic and hence the decoherence rate is the
same for the two N\'eel states. Note that Eq.(\ref{pilot})
has been written in the N\'eel-state basis, implicitly assuming the
decoherence term to be much larger than the effective coupling, $V$.

The resulting set of equations can be easily solved by defining the population difference:
$\Delta {P}= \rho_{11} - \rho_{22}.$
The solution is 
\begin{equation}
\Delta {P}=\frac{-4 V^2}{\sqrt{\Gamma^2-16V^2}}  (\frac{e^{\omega_+ t}}{\omega_+} -
\frac{e^{\omega_- t}}{\omega_-}),
\label{Delta}
\end{equation}
 when initially the chain is {in the $|N_1\rangle$ state.} 
The values of $\omega_{\pm}$ are $  0.5 \,
( -\Gamma \pm \sqrt{\Gamma^2 - 16 V^2})$.  Then, for long times,
Eq. (\ref{Delta}) leads to equal populations of the two N\'eel states,
$\rho_{11}=\rho_{22}$. 
{In parallel, non-diagonal terms (coherences) die out on the same time
scale, so that the final state of the evolution is an equal and incoherent
population of the two N\'eel states.}

In the present case, the decoherence rate is very fast compared to the
Rabi evolution, $\Gamma \gg V$ (or equivalently $T_{Rabi} \gg T_2^*$). Therefore, the population difference can
be approximated by:
\begin{equation}
\Delta {P} \approx e^{-\frac{4 V^2 t}{\hbar \Gamma} },
\label{Delta2}
\end{equation}
{from which we} 
obtain the \textit{switching rate} between N\'eel {states  $|N_1\rangle$ and $|N_2\rangle$:}  
\begin{equation}
\frac{1}{\tau_{1\rightarrow 2}} = \frac{2 V^2}{\hbar \Gamma} =  \frac{1}{ T_{Rabi}} \frac{2 \pi V}{ \Gamma}
\label{switching}
\end{equation}

The decoherence rate, $\Gamma/\hbar$, accounts for the loss of coherence
of the evolving spin state due to  collisions with hot electrons from
the substrate. Its calculation is akin to the calculation of de-excitation of spin
states on surfaces~\cite{Novaes2010,PSS} {except that the spin state
does not change.  Each collision entails a change} in the state phase
and hence the decoherence of the evolving spin state, $|N_1\rangle$
or {$|N_2\rangle$}. The decoherence rate can  be expressed as
the addition of electron collisions with each atom of the
chain. As shown in Ref.~\cite{Novaes2010}, the electron collision rate
 can be separated in three parts, and the total
 rate can be expressed as 
\begin{equation} 
\Gamma=N T_{Fe}(E_F) \frac{k_B
T}{2\pi} P_{Spin} (N_1 \rightarrow N_1).  
\label{rate} 
\end{equation}
 The first {factor} is the electron transmission at the Fermi energy, $N
T_{Fe}(E_F)$, that takes into account the {electron} flux through the $N$
Fe atoms of the chain, where $T_{Fe}(E_F)${ typically amounts to
1~\cite{Novaes2010}~\footnote{Recent calculations by~\cite{Novaes2014} on
individual Fe adatoms on Cu$_2$N/Cu(100) yielded $T_{Fe}(E_F)$ values in
the same range as those obtained for Mn adsorbates in~\cite{Novaes2010}.
}}. The second factor accounts for the
number of electrons that can collide with the target. These hot
electrons are given by the convolution of the electron and hole Fermi
functions~\cite{Jaklevic}. At low temperature, $T$, the second term
is just $\frac{k_B T}{2\pi}$. { Finally, $P_{Spin}(N_1 \rightarrow
N_1)$ represents the elastic scattering probability for an electron
scattered by an atom in the chain; it is not a trivial number since
it depends on all the possible spin transitions of the target. It
amounts typically to 0.8-0.9 for the various chains (see a discussion
in ~\cite{Gauyacq2013a}).} 

\begin{figure}
\includegraphics[angle=-90,width=0.52\textwidth]{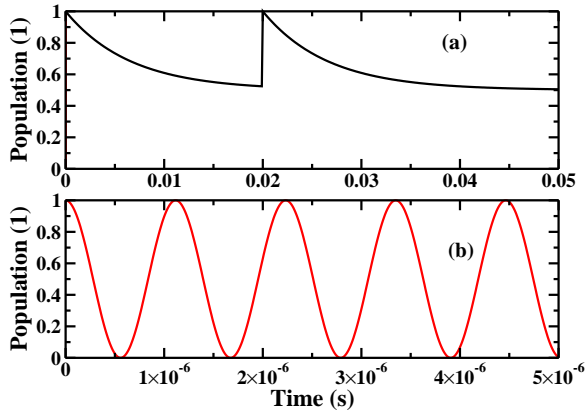}
\caption{
$(a)$ Time evolution of the N\'eel-state population in the presence
of decoherence.  A {Fe$_6$} chain is initially prepared
in the $|N_1\rangle$ state, and decoherence leads to an exponential
decline of its population.  At $\sim 0.02$ s, the chain is either in
the $|N_1\rangle$ or the $|N_2\rangle$ states with equal probability.
We assume that a measurement is performed at $\sim 0.02$ s in order to
determine the chain's state, and hence the population is determined and
the evolution starts again.  $(b)$ For comparison, the $|N_1\rangle$-state
population is shown in the absence of decoherence. The evolution is of
the Rabi type and the oscillation { is orders of magnitude
faster than when  decoherence is taken into account as in $(a)$ (note
the change in time scale).}
 \label{figure2}}
\end{figure}

{F}or a six-atom Fe chain on Cu$_2$N/Cu(100),
the decoherence rate is indeed very fast.  The effective coupling,
$V$, is $1.85 \times 10^{-6}$ meV, while the decoherence rate
 is 0.14 meV at $T=2$~K.  {Thus,
the decoherence rate is much larger than the Rabi frequency between
the N\'eel states, justifying the above approach.  The consequence
of fast decoherence} is clearly seen in Fig. \ref{figure2}. In $(a)$,  the
evolution of the population of state $|N_1\rangle$ {in the presence
of decoherence} is shown as a function of time,  {for an initial population of $|N_1\rangle$
equal to 1}. It decreases exponentially to 0.5, and the system becomes an
incoherent superposition of states 1 and 2. Figure~\ref{figure2}~(a)
also illustrates the effect of a measurement as performed e.g. in an STM
experiment: at an arbitrarily chosen
time ($t=0.02$ s), a measurement is performed that{, for example,} finds the system in
state $|N_1\rangle$; this leads to a restart of the population evolution
that again relaxes towards an incoherent situation.

 In Fig.~\ref{figure2}~$(b)$ the Rabi evolution is shown
for comparison. The time evolution is orders of magnitude
slower in the presence of decoherence
(see e.g. Eq.(~\ref{switching})).  {The  quantal state
of the chain loses its phase many times before a Rabi oscillation could
be performed (before quantum tunnelling could set in) and the decoherence strongly
hampers the switching rate between the two spin states.} {
The difference between cases (a) and (b) goes beyond the difference in
time scales. Rabi oscillations (b) correspond to a reversible coherent
evolution between the two N\'eel states. Decoherence results in an
irreversible evolution toward an incoherent superposition of the two
N\'eel states and thus explains the experimental observation by Loth
\textit{et al.}~\cite{Loth2012} of N\'eel states and not of
the ground state of the {isolated} system.} 
{However,
decoherence is not the only physical effect at play, the relative
values of magnetic anisotropy ($D$ and $E$) and exchange couplings ($J$) are very
important. Indeed,  N\'eel states have not been observed for chains with
an even number of  Mn atoms~\cite{Heinrich2006} due to the negligible
longitudinal anisotropy, $D$ in Eq.~(\ref{hamiltonien}).}

\begin{figure}
\includegraphics[angle=-90, width=0.50\textwidth]{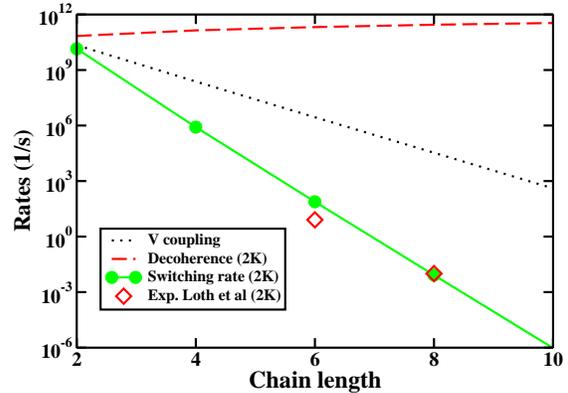}
\caption{{Rates for the various processes in 
the Fe chains as a function
of the number of atoms for a substrate temperature equal to 2K. }The available experimental data (red diamonds) for the switching rate are also shown~\cite{Loth2012}.
The full green line is the present computed switching rate between the two N\'eel states. The $V$ coupling and and the decoherence rate are  plotted as dotted and dashed lines, respectively.
 \label{figure3}}
\end{figure}

{The switching rate between the two N\'eel
  states, $1/\tau_{1\rightarrow 2}$, obtained in our calculations
  (Eq. (\ref{switching})), is displayed in Figure \ref{figure3}. It }
  shows the different rates as functions of the number
of atoms in the chain. The switching rate presents a {fast} exponential
decrease with the number of atoms in {quantitative}
agreement with the switching rates measured  for {Fe$_6$} and {Fe$_8$}
\cite{Loth2012}. The exponential decrease with increasing length is
due to the exponential behavior of the Rabi frequency, also shown in
Fig. \ref{figure3}. {Indeed, when increasing the chain
length, the indirect coupling between N\'eel states becomes of higher
and higher order {in $J$, Eq.~(\ref{hamiltonien})}, leading to
the exponential decrease of the effective coupling, $V$, {with
the number of atoms, $N$}.} However, the
decoherence rate slightly increases with chain {length,
due to the increase of the number of atoms, $N$, that can be hit by substrate
electrons, {Eq.~(\ref{rate})}.}

Our calculations show that the switching rate presents \textit{three} regimes
as a function of  temperature. $(i)$ A high-temperature regime where hot substrate
electrons {have enough energy to excite higher-lying
excited states in the chain, trigering an efficient indirect switching
process, such as {the one}} described in Refs.~\cite{Loth2012,Gauyacq2013a}.
$(ii)$
A second regime of intermediate temperatures where indirect
transitions are not possible due to energy conservation. This second
regime{, studied in detail in the present work,}
is dominated by hot-electron decoherence and present a $\sim 1/T$
behavior. In front of the
exponential behavior of the high-$T$ regime, this $1/T$ is slowly
varying and can be easily confused with a tunneling behavior. $(iii)$ Finally,
at {extremely}  low temperatures, decoherence becomes
negligible and Rabi oscillations dominate. In
 this very-low-temperature regime,  {the chain's evolution
can be described via  Hamiltonian (\ref{hamiltonien}),  and truly
corresponds to a quantum tunneling regime. In this regime, a decay term
for the excited state, $|EXC\rangle$, population has to be included.
For the six-atom chain, the decoherence rate becomes equal to the Rabi
frequency  in the $\mu$K range  and as the chain length increases, the
cross-over from decoherence-dominated to quantum tunneling behavior
is rejected to further lower temperatures.}

{As the chain length shortens, the importance of
decoherence  in the spin dynamics reduces.}
{The situation of  {Fe} dimers is different
from that of longer chains. At 2~K, (see Fig.~\ref{figure3}),
 all rates are roughly of the same order of magnitude, so that
the system is not decoherence-dominated. {This is
the case of Fe dimers at 330 mK~\cite{Bryant2013}, where} the decoherence rate
is smaller than the coupling $V$ and consistently, the conductance
spectrum shows a low-energy threshold due to the excitation of
the $|EXC\rangle$ state.}

In summary, we have evaluated the spontaneous switching rate between
the two possible N\'eel states of atomic Fe chains adsorbed on
Cu$_2$N/Cu(100)~\cite{Loth2012}. We show that the quantum tunneling
associated with Rabi oscillations in this system is {deeply modified
by the presence of a strong substrate-induced decoherence process,
leading to a considerable slowing down of the switching process and
to a change of its character. The decoherence-assisted process leads
to the relaxation of the chain magnetic state toward an incoherent
population of both N\'eel states, instead of  the Rabi oscillations
of the quantum tunneling process.} {At high temperature, indirect
transitions~\cite{Gauyacq2013a} induced by hot substrate electrons
lead to an  Arrhenius-type of $T$-dependence~\cite{Loth2012}. In the
few-Kelvin range, the decoherence-assisted switching mechanism takes
over,  in quantitative agreement with the experiment, and {with} the
correct scaling with chain length, the switching {rate} slowing down
exponentially with the chain length.}
 Only at extremely low temperatures ({below the $\mu$K range} for
the six-atom chain and exponentially lower for longer chains) does
quantum tunneling prevail over the above decoherence-assisted
mechanism. 

Decoherence appears as a key phenomenon in the
switching of
magnetization at surfaces, and it is quite remarkable that
decoherence is able to dominate the spin dynamics {down
to unattainable temperatures}
even for a spin system of only a few atoms.

N.L. acknowledges financial support from Spanish MINECO (Grant
No. MAT2012-38318-C03-02 with joint financing by FEDER Funds from the
European Union) and fruitful discussions with J. Fern\'andez-Rossier.

\bibliography{references}
\end{document}